\documentclass[10pt,twocolumn]{article}

\usepackage{graphicx}
\usepackage{dcolumn}
\usepackage{bm}
\usepackage{array}
\usepackage{booktabs}
\usepackage{multirow}
\usepackage{lipsum}
\newcommand{\head}[1]{\textnormal{\textbf{#1}}}

\usepackage{xcolor}
\usepackage{graphicx}
\usepackage{graphics}
\usepackage{amsmath}
\usepackage{amssymb}
\usepackage{amsthm,amstext}
\usepackage{amsfonts}
\usepackage{mathrsfs}
\usepackage{MnSymbol}
\usepackage{mathrsfs}
\usepackage{stmaryrd}
\usepackage{latexsym}
\usepackage{bm}
\usepackage[utf8]{inputenc}
\usepackage{mdframed}
\usepackage{hyperref}
\usepackage{lipsum}
\usepackage{geometry} 
\usepackage{xtab,afterpage}
\usepackage{lipsum} 

\newcommand\bx{\textbf{\emph{x}}}

\newcommand\bi{\textbf{\emph{i}}}

\newcommand\F{\textbf{F}}

\newcommand\G{\textbf{G}}

\newcommand\sx{\textbf{\textsf{x}}}

\renewcommand\d\delta
\newcommand\D\Delta
\newcommand\e{\varepsilon}
\newcommand\s{\sigma}
\newcommand\ph{\varphi}

\newcommand\bxi{\boldsymbol{\xi}}

\newcommand\Div{\text{Div}}

\newcommand\scrW{\mathscr{W}}

\newcommand\bbR{\mathbb{R}}

\usepackage{etoolbox}
\makeatletter
\patchcmd{\@maketitle}{\begin{center}}{\begin{flushleft}}{}{}
\patchcmd{\@maketitle}{\begin{tabular}[t]{c}}{\begin{tabular}[t]{@{}l}}{}{}
\patchcmd{\@maketitle}{\end{center}}{\end{flushleft}}{}{}
\makeatother

\newcommand\beq{\begin{equation}}
\newcommand\beqn{\begin{eqnarray}}
\newcommand\eeq{\end{equation}}
\newcommand\eeqn{\end{eqnarray}}
\newcommand{\bbar}{\overline}

\newcommand{\trasp}{\hspace{-1pt}^\textsf{T}}

\begin{document}

\title{\textbf{Catastrophic thinning of dielectric elastomers}}



\author{Zurlo G., Destrade M.\\
School of Mathematics, Statistics and Applied Mathematics, \\
NUI Galway, University Road, Galway, Ireland;\\[10pt]
DeTommasi D., Puglisi G.\\
Dipartimento di Scienze dell' Ingegneria Civile e dell' Architettura, \\
Politecnico di Bari, Via Re David 200, 70125 Bari, Italy.}

\date{\today}

\maketitle

\textbf{
We provide a clear energetic insight into the catastrophic nature of the so-called creasing and pull-in instabilities in soft electro-active elastomers. 
These phenomena are ubiquitous for thin electro-elastic plates and are a major obstacle to the development of giant actuators;
yet they are not completely understood nor modelled accurately. 
Here, in complete agreement with experiments, we give a simple formula to predict the voltage thresholds for these instability patterns and model their shape, and show that equilibrium is impossible beyond their onset.  Our analysis is fully analytical, does not require finite element simulations, and can be extended to include pre-stretch and to encompass any material behaviour.
}
\\[10pt]
Consider a thin dielectric plate with conducting faces: when will it break if a voltage is applied? If it is rigid it will break once its dielectric strength is overcome by the voltage. But what if it is highly stretchable, like the elastomers used for soft actuators, stretchable electronics, artificial muscles or energy harvesters? The precise answer to that question is not known. Experiments show that it will break when highly localised thinning deformations occur, and further, that this process is \emph{catastrophic}, in the sense that the deformations cannot be controlled or restrained once they have started. 

Here we unveil the physical meaning of catastrophic thinning, based on energy minimisation arguments; 
we derive a unifying and simple formula giving very accurate predictions of the voltage thresholds for the \emph{creasing instability} (one-side constrained plates, with one compliant electrode on one face and a rigid electrode on the other face) and the \emph{pull-in instability} (unconstrained plates with fully compliant electrodes glued on both faces);
we calculate the shape of the instability patterns at their onset and show that they are not sustainable but give the path to final breakdown; 
we generalise the results to include a pre-stretch.

There are several protocols in place to deform thin dielectric elastomers by  applying a voltage. 
Typically they are realised with highly deformable (isotropic, incompressible) polymeric or silicone electro-elastic films, brushed with conductive carbon grease \cite{Pelr00}. 
With an electric field the attractive Coulomb forces between the electrodes compress the film along the thickness direction, which then expands in the planar direction. 

\begin{figure}[!h]
\centering
\includegraphics[scale=.48]{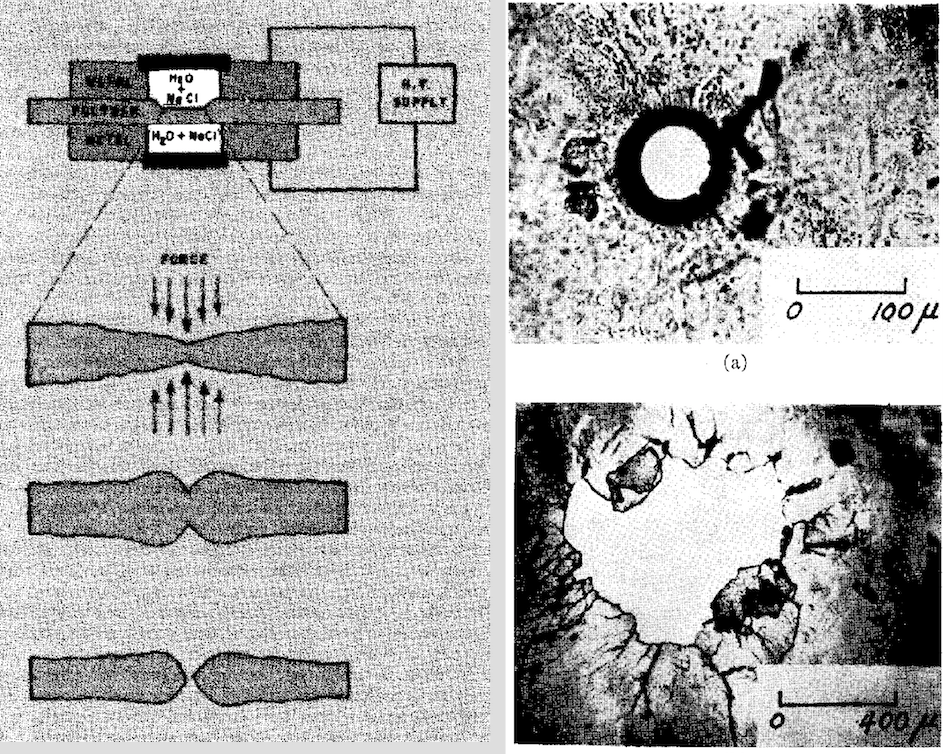}
\caption{\footnotesize{\label{BlokLeGrand}} Right: First recorded experimental evidence of catastrophic localisation of thinning deformations in unconstrained films (pull-in instability), by Blok and LeGrand\cite{BlokLeGrand}, 1969. 
Left: Sketch of the experimental setting and qualitative description of the onset of localisations.}
\end{figure}

At first the rectangular film deforms into another rectangle and the deformation remains homogeneous until a \emph{critical voltage} is reached. 
Then, a sudden, irreversible and non-homogeneous thinning localisation occurs (usually accompanied by a spark and a popping sound \cite{Suo12}), anticipating the dielectric breakdown of the film, forming holes that affect its insulating power, and significantly reducing its capacitance (Fig.\ref{BlokLeGrand}). 

Call $E$ the following non-dimensional measure \cite{Suo1} of the electric field 
\begin{equation}\label{main}
E = \sqrt{\dfrac{\epsilon}{\mu}} \dfrac{V}{h},
\end{equation}
and $E_c$ its critical value.
Here $V$ is the applied voltage, $\epsilon$ is the dielectric permittivity of the elastomer, $\mu$ its initial shear modulus and $h$ its initial thickness.
For unconstrained plates, experiments \cite{Pelr98} give $E_c$ in the range $0.678-0.686$ at the onset of pull-in instability, which corresponds to a contraction of $30-34\%$ in the layer's thickness, depending on the composition of the elastomer. 
For one-side constrained plates, experimental measurements \cite{WaEE11} reveal that $E_c \simeq 0.85$ at the onset of creasing instability.
Finally, extensive experimental studies document that dead-loads (for unconstrained films) and pre-stretch (for one-side constrained films) play a beneficial role in delaying the onset of instabilities \cite{Suo12b, WangCreasingPrestretch11, WangCreasingPrestretch}. 

However, despite the abundance of experimental data and the pressing demand for reliable models, so far the critical values of the electric field have not been predicted by the theory in an entirely satisfying and accessible manner. 
Many papers establish a connection between pull-in and snap-through instabilities for unconstrained films \cite{Suo1} using the Hessian method, but when dead-loads are applied to the layer, these predictions fail to account for the actual delay of the onset of instability, a key factor for technological applications \cite{Suo12b}. 
Indeed, dead-loads can suppress snap-through instability but not the catastrophic localisation of deformations leading the device to failure. 
For unconstrained films there have been attempts at introducing linearised \cite{BeGe11,DePE13,DoOg14,Zu} and non-linear \cite{DePE13b} non-homogeneous bifurcation modes on top of the homogeneous deformation, but they require lengthy calculations, do not explain the catastrophic nature of localised deformations for unconstrained films and do not quantify the beneficial effects of pre-stretch.
The situation is even worse for one-side constrained films. 
So far, electro-creasing has only been studied with entirely numerical methods based on finite element method (FEM) simulations: in the absence of pre-stretch they lead to an estimate of $E_c \simeq 1.03$, which is more than 20\% off the experimental mark \cite{WaEE11} of $E_c \simeq 0.85$. 
There are no theoretical predictions available when pre-stretch is applied.

\begin{figure}[!th] 
\begin{centering}
\includegraphics[width=0.5\textwidth]{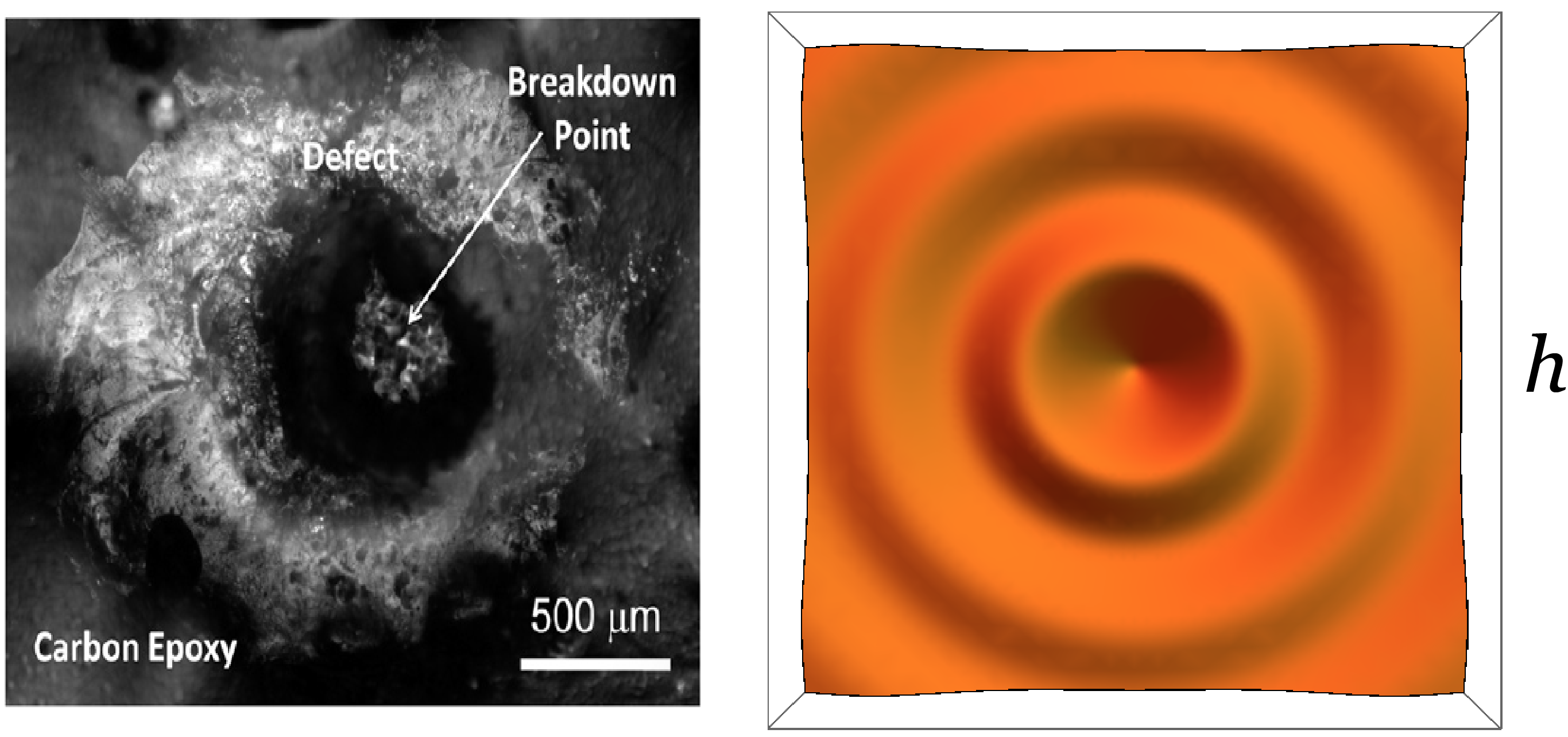}
\caption{\label{Energy1}
\footnotesize{Left: Experimental creasing instability for non pre-stretched one-side constrained films (taken from \cite{Zhang11}). 
Right: Predicted shape of the failure precursor in films of thickness $h$.}}
\end{centering}
\end{figure}

\begin{figure}[!th] 
\begin{centering}
\includegraphics[width=0.4\textwidth]{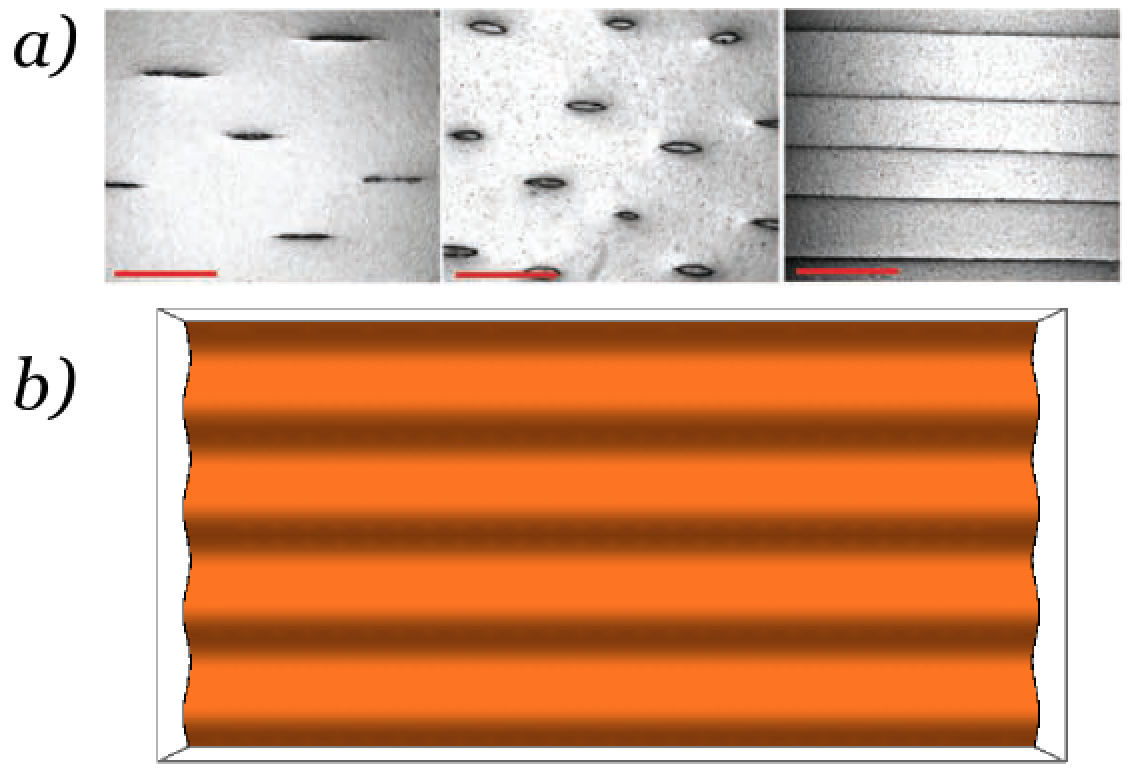}
\caption{\label{Energy2}
\footnotesize{a) Alignment of creases in the direction of higher pre-stretch as voltage increases for one-side constrained films (taken from \cite{WangCreasingPrestretch}). { b)} Failure precursors in films with thickness $h$ and pre-stretch $\lambda_1=2$, $\lambda_2=1$.}}
\end{centering}
\end{figure}

Our analysis (discussed below and detailed in the Supplementary Material (SM) section) provides a new paradigm for understanding electromechanical instability, which we find corresponds to a threshold where the electro-elastic energy does not possess minimisers in a general class of homogeneous and non-homogeneous deformations. For both unconstrained and constrained films, with and without pre-stretch and for a quite general class of incompressible materials, we obtain the following simple \emph{unifying formula} for the critical electric field: 
\beq\label{EgenEcr}
E_c = \frac{2}{\sqrt{3}}\sqrt{\frac{W'(I)}{\mu}}\min\left(\frac{1}{\lambda_1},\frac{1}{\lambda_2}\right)
\eeq
where $\lambda_1$ and $\lambda_2$ are the principal stretches in the plane of the thin layer, $W$ is the elastic energy density of the incompressible dielectric film and $I=\lambda_1^2+\lambda_2^2+(\lambda_1\lambda_2)^{-2}$ is the first invariant of deformation 
(and hence, $\mu =2 W'(3)$). For unconstrained films, the principal stretches are determined by the homogeneous solution; for constrained films, they are the fixed pre-stretches imposed prior to attachment to the rigid substrate.
Above the critical electric field no stable configurations exist, neither homogeneous {\it nor} non-homogeneous. 
As soon as $E_c$ is attained, failure precursors appear, opening the way for a catastrophic failure of the film. The initial pattern of these precursors is a permanent signature for the subsequent inelastic processes, see Figs. \ref{Energy1}, \ref{Energy2}.

\begin{figure}[!th] 
\begin{centering}
\includegraphics[width=0.4\textwidth]{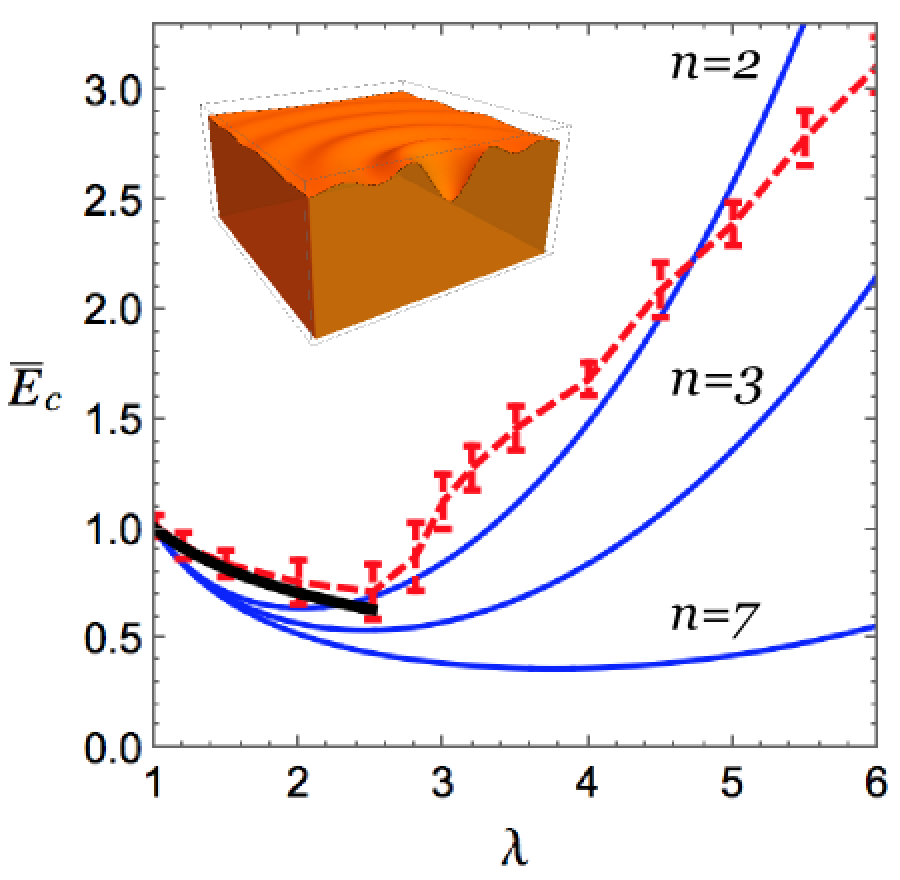}
\caption{\label{CreasingFig}
\footnotesize{Creasing instability in the presence of uniaxial pre-stretch; comparison of experiments (dashed line with error bars) \cite{WangCreasingPrestretch} and theory (blue solid curves, drawn for different values of $n$, the number of links in the 8-chain Arruda-Boyce model).
Here $\overline E_c=E_c/\sqrt{2/3}$, see SM for details on the model calibration. Black solid curve: previous qualitative trend based on dimensional/finite element analysis \cite{WangCreasingPrestretch}.
}}
\end{centering}
\end{figure}

To assess the capability of formula \eqref{EgenEcr} to predict the onset of catastrophic thinning we first consider one-side constrained films for the \emph{creasing instability}, see sketch in the insert of Fig.\ref{CreasingFig}.
\begin{table*}
\centering
\begin{tabular}{lcccccccccc}
\toprule[1.5pt]
& \multicolumn{2}{c}{\head{Creasing}} & \multicolumn{6}{c}{\head{Pull-in}}\\
\cmidrule(lr){2-4}\cmidrule(l){5-9}
\multirow{1}{*}{\small Pre-stretch (dead-load)} 
& $1$ & $1-2.5$ & $2.5-6$ & $1$ & (20\,g) & (25.5\,g) & (31\,g) & (36.5\,g)\\
\cmidrule(lr){2-4}\cmidrule(l){5-9}
\multirow{2}{*}{\small $E_c$ (or $V_c$, in kV) experiments} 
& $0.85$ & $\boldsymbol{\checkmark}$ & $\boldsymbol{\checkmark}$ & $0.678\hspace{-2pt}-\hspace{-2pt}0.686$ & $ \boldsymbol\checkmark$ &$\boldsymbol\checkmark$ & $\boldsymbol\checkmark$ & $\boldsymbol\checkmark$ \\
& \cite{WaEE11} & \cite{WangCreasingPrestretch} & \cite{WangCreasingPrestretch} & \cite{Pelr98}  & \cite{Suo12b} & \cite{Suo12b} & \cite{Suo12b} & \cite{Suo12b}  \\
\cmidrule(lr){2-4}\cmidrule(l){5-9}
\multirow{2}{*}{\small $E_c$ (or $V_c$, in kV) other theories} 
& $1.03$  & $\text{\sffamily x}$ & $\text{\sffamily x}$ & $0.687$ & $\boldsymbol\checkmark$ & $\text{\sffamily x}$ & $\text{\sffamily x}$ & $\text{\sffamily x}$ \\
& FEM\cite{WaEE11} & dim$\dagger$\cite{WangCreasingPrestretch} & $\text{\sffamily x}$ & H$\ddagger$\cite{Suo1} & H\cite{Suo12b} & $\text{\sffamily x}$ & $\text{\sffamily x}$ & $\text{\sffamily x}$ \\
\cmidrule(lr){2-4}\cmidrule(l){5-9}
\multirow{1}{*}{\small $E_c$ (or $V_c$, in kV)  our theory} 
& $0.816$ & Fig.\ref{CreasingFig} & Fig.\ref{CreasingFig} & 0.680 & Fig.\ref{PullinFig} & Fig.\ref{PullinFig} & Fig.\ref{PullinFig} & Fig.\ref{PullinFig} \\
\bottomrule[1.5pt]
\end{tabular}
\caption{Experimental results Vs theoretical predictions of previous models and present model, which leads to the simple formula \eqref{EgenEcr}. $\dagger$Only a qualitative behaviour was obtained in \cite{WangCreasingPrestretch} for this stretch range, based on heuristic dimensional analysis.
$\ddagger$Hessian method.}
\end{table*}

For small stretches ($\lambda_{1}, \lambda_{2} \le 1.5$) the constitutive response of the soft dielectric polymer can be modelled as a neo-Hookean solid, for which $W=\mu(I-3)/2$ and the formula \eqref{EgenEcr} further simplifies to $E_c=\sqrt{2/3}\,\min(1/\lambda_1,1/\lambda_2)$. 
In the absence of pre-stretch ($\lambda_1=\lambda_2=1$) this formula gives $E_c=\sqrt{2/3} = 0.816$, less than 4\% off the  value $E_c \simeq 0.85$ obtained experimentally (\cite{WaEE11}, Fig 3b) and closer than the estimate $E_c = 1.03$ obtained by FEM simulations (\cite{WaEE11}, Fig 4b). 
For small values of uniaxial pre-stretch ($\lambda_1=\lambda>1,\, \lambda_2=1$), experiments \cite{WangCreasingPrestretch} report an initial reduction of the critical electric field, which is in agreement with the prediction of formula \eqref{EgenEcr} for the neo-Hookean model, $E_c=\sqrt{2/3}\,\lambda^{-1}$. Beyond this initial negative effect, larger values of pre-stretch were measured as beneficial in increasing the critical electric field \cite{WangCreasingPrestretch11, WangCreasingPrestretch}, an effect that did not receive theoretical or numerical explanations so far. 
To demonstrate the ability of formula \eqref{EgenEcr} in reproducing this effect, we use a material model $W$ which accounts for the strain-stiffening induced by the limit chain extensibility of the elastomer in large deformations. 
Fig. \ref{CreasingFig} shows the good agreement reached between our theory and experiments (see SM for details on the calibration of the model). 

\begin{figure}[!th] 
\begin{centering}
\includegraphics[width=0.4\textwidth]{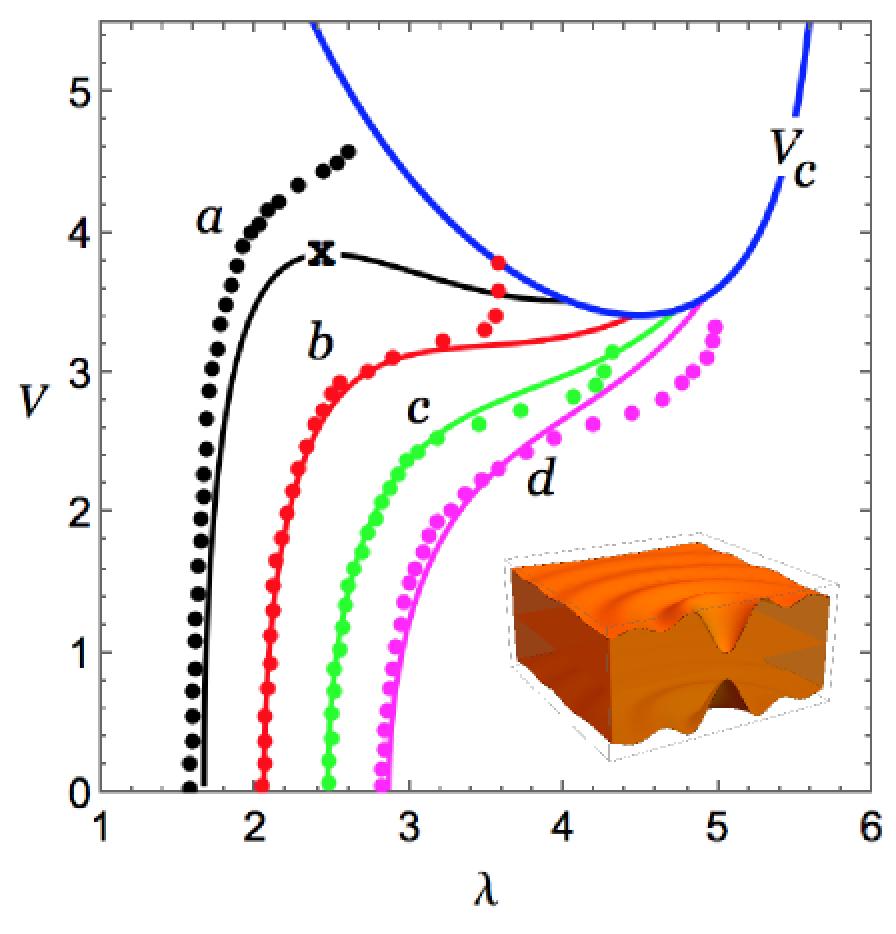}
\caption{\label{PullinFig}
\footnotesize{Pull-in instability in equi-biaxially strained films pre-stretched by equal dead-loads; comparison of experimental (dots) \cite{Suo12b} and theoretical (solid curves) critical voltages (in kV). 
Curves $a,b,c,d$ correspond to dead-loads weighing $20$g, $25.5$g, $31$g, $35.6$g, respectively. 
Dots describe homogeneous paths until failure occurs. 
Solid curves represent the homogeneous loading paths, tuned to reproduce the dotted paths. 
Their intersection with the dashed curve (given by Eq.\eqref{EgenEcr}) corresponds to catastrophic thinning (see details on the calibration in the SM). 
The $\sx$ on the $a$ curve denotes failure according to the Hessian approach,  coinciding with loss of monotonicity of the loading curve. 
Observe that for higher dead-loads the loading curves are always increasing and the Hessian approach cannot predict failure.
}}
\end{centering}
\end{figure}

We then consider unconstrained films to study the onset of the \emph{pull-in instability}, see sketch in the insert of Fig.\ref{PullinFig}. 
For equi-biaxially stretched films  ($\lambda_1=\lambda_2=\lambda$) in the absence of dead-loads, the homogeneous extension of the neo-Hookean elastomer is described by (\cite{Suo1}, see also SM)  $E=\sqrt{\lambda^{-2} - \lambda^{-8}}$. 
It reaches $E=E_c = \sqrt{2/3}\lambda^{-1}$ when \cite{DePE13} $\lambda = 3^{1/6}$ and then, $E_c = \sqrt{2}/3^{2/3}  = 0.680$, falling squarely within the range of the experimental values $0.678-0.686$ in the absence of pre-stretch \cite{Pelr98}. 
Thus, in absence of dead-loads, our analysis is close to the Hessian approach, which gives an estimate of $E_c =0.687$ \cite{Suo1}.

Things change drastically when the film is pre-stretched by applying dead-loads prior to the voltage; then experiments show that  {\it giant areal gains} can be achieved \cite{Suo12b, Jiang16, BoLi}. 
Theoretical models based on the Hessian approach fail to predict the actual gain, in particular for higher dead-loads. 
That's because the Hessian criterion detects the points where the voltage-stretch curve ceases to be increasing \cite{Suo1}. 
But elastomers stiffen greatly at large strains, and sufficiently high dead-loads will make the voltage-stretch curve monotonic increasing: this phenomenon can lead to the erroneous conclusion that electromechanical instability can be eliminated by high pre-stretch, in contradiction with experiments \cite{Suo12b, Roentgen}.
Consider for example experiments on a voltage actuated silicone disk, pre-stretched by dead-loads, as carefully conducted and described in \cite{Suo12b}. 
Based on the experimental results reported for the purely mechanical behaviour of VHB silicone, we may obtain the homogeneous loading curves for different values of dead-loads using a strain-stiffening model (see SM for details). 
For lower values of pre-stretch (see Fig.\ref{PullinFig}), the voltage-stretch curves have a peak, corresponding to failure according to the Hessian condition, but this peak disappears for higher values of pre-stretch and the Hessian condition is no longer violated. 
Nonetheless, the experimental plots have a maximum, clearly corresponding to failure, see last upper experimental dots in Fig.\ref{PullinFig}.
Our theory predicts failure whenever the homogeneous loading curves intersect the critical threshold curve described by formula \eqref{EgenEcr}. 
It gives a clear correspondence between experimental and theoretical thresholds, as seen in  Fig.\ref{PullinFig}.

We may thus conclude that the simple formula \eqref{EgenEcr}, coupled to that based on snap-through (Hessian) modelling \cite{Norr08, Suo1}, provides a complete picture of the electric breakdown experienced by unconstrained and one-side constrained voltage-actuated thin dielectrics, see summary in Table 1.

\begin{figure}[!th] 
\begin{centering}
\includegraphics[width=0.4\textwidth]{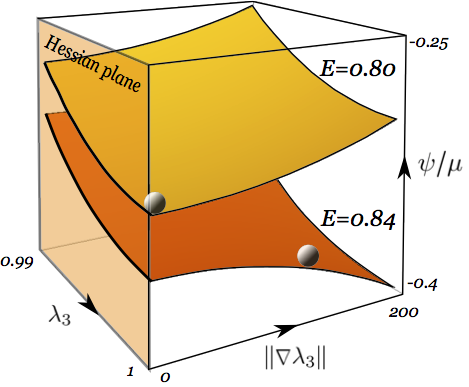}
\caption{\label{Fig-Energy}
\footnotesize{
Electro-elastic free energy $\psi/\mu$ for a one-side constrained film without pre-stretch ($\lambda_1=\lambda_2=1)$. 
Here the film is modeled as a slightly compressible neo-Hookean material of thickness $h=0.01L$ where $L$ is a typical lateral length scale and $\kappa/\mu=1000$ where $\kappa$ is the initial bulk modulus (see SM (7),(8)). 
For $E<E_c=0.816$ the homogeneous solution with $\lambda_3=1$ is the unique energy minimiser, see yellow surface. 
Above $E_c$ the energy is still convex in $\lambda_3$, but becomes non-convex in $\nabla\lambda_3$, see orange surface: failure precursors are then energetically favoured but no energy minimisers exist. 
In the shaded {\it Hessian plane} (at $||\nabla\lambda_3||=0$), Hessian stability takes place; there the energy is convex in $\lambda_3$ and the failure threshold cannot be captured.}}
\end{centering}
\end{figure}

Our theory is based on an energy minimisation argument \cite{DePE13b}. 
{\it Stable} equilibrium configurations are minimisers of the total electro-elastic free energy $\Psi=U-QV/2$, where $U$ is the total elastic energy, $Q$ the total charge on the electrodes and $V$ the applied voltage (see SM for details), while the surface energy of the polymer layer is negligible in the present context \cite{WangCreasingPrestretch}. 
Under the assumption that the membrane is thin and that curvature effects can be neglected \cite{PZ}, the electric field in the film can be approximated by the ratio $V/(h \lambda_3)$, where $\lambda_3$ is the (possibly non-homogeneous) thickness stretch. 
With these assumptions, the total free energy depends only on the film deformation. 

The thinness of the dielectric film allows us to perform a Taylor expansion of $\Psi$ in powers of $h$ and truncating $o(h^3)$ terms gives the minimal tools for detecting the onset of localisation instabilities (higher-order terms are required for post-critical analysis, which is beyond the scope of this work.) 
Denote by $S$ the constrained (lower) surface for creasing, and the mid-surface for pull-in.
Then we obtain $\Psi=\int_S\psi\,da$ where the surface energy density $\psi$ is expanded as 
\beq
\label{energycreasing}
\psi(\lambda_i,\nabla\lambda_3) = h\,\ph(\lambda_i) + h^3 \left(\alpha_1(\lambda_i)\lambda_{3,1}^2 + \alpha_2(\lambda_i)\lambda_{3,2}^2\right).
\eeq
Here the stretches $\lambda_i$ are functions of the planar coordinates $(x_1,x_2)$ and $\nabla\lambda_3 = (\lambda_{3,1},\lambda_{3,2})$ is the gradient of thickness stretch, which explicitly accounts for non-homogeneity of deformation modes. 
While the first-order energy term $\ph$ only depends on local measures of deformation, the higher-order terms, which play a major role in our theory, weigh on the energetic cost to develop non-homogeneous deformations. 

When higher order terms are not considered, only homogeneous configurations can be described. 
Homogeneous stationary configurations are characterised by the Euler-Lagrange equations $\partial \ph / \partial \lambda_i = 0$ and their stability is assessed through convexity of $\ph$: for electroelastic systems this is the so-called \emph{Hessian approach} \cite{Norr08, Suo1}.  
When applied to pull-in  this approach cannot predict instability neither for non-equibiaxial states of deformation (see SM) nor for large deformations (see above and  \cite{Suo12b}). 
When applied to creasing (for undeformed or equi-biaxially pre-stretched films) it predicts that the undeformed configuration is stable for any voltage, which is clearly contradicted by experiments. 

These shortcomings can be addressed by taking in consideration the whole energy \eqref{energycreasing}, which imposes more stringent requirements for the existence of minimisers (see SM for details). 
Indeed, direct methods of calculus of variations dictate that minimisers of $\Psi$ exist provided that $\psi$ is convex in $\nabla\lambda_3$.
In terms of the energy \eqref{energycreasing}, convexity in $\nabla\lambda_3$ means that the functions $\alpha_{1}$, $\alpha_2$ are both positive, and this is the case as long as the electric field is less than the critical threshold $E_c$ defined by \eqref{EgenEcr}. 

Above $E_c$ the total free energy becomes concave in $\nabla \lambda_3$ and no energy minimisers exist (be they homogeneous or belonging to a wide class of non-homogeneous ones). 
Nonetheless, immediately above $E_c$, non-homogeneous {\it failure precursors} of deformation become possible: as soon as they appear, they are energetically more favourable than the homogeneous state, since they can lower the total free energy, albeit without finding a minimum (Fig.\ref{Fig-Energy}). This explains the {\it catastrophic nature} of localised thinning in electroelastic films. 
Experiments on one-side constrained films show that when the film is not pre-stretched, localisation mainly takes place in circular spots, whereas when the film is pre-stretched in one direction prior to bonding to the rigid substrate, the resulting thinning localisations tend to align in the direction of higher pre-stretch \cite{WangCreasingPrestretch}, see Figs.\ref{Energy1},\ref{Energy2}.
Both findings are easily covered by our theory. 

Indeed, linearising the Euler-Lagrange equation based on \eqref{energycreasing}, we find that failure precursors of the type $\lambda_3(x_1,x_2) = \lambda_3^o + w(x_1,x_2)$, with $\lambda_3^o$ a constant  and $w$ small, solve (SM for details)
\beq\label{www}
\alpha_1 w_{,11} + \alpha_2 w_{,22} - \Gamma\,w = 0, 
\eeq
where $\alpha_{1}$, $\alpha_2$ and $\Gamma$ are functions of the underlying stretch and electric field. 
For almost incompressible materials $\Gamma$ is always positive whereas, as we have seen, $\alpha_{1}$, $\alpha_2$ become negative above the critical voltage defined by \eqref{EgenEcr}. 
For undeformed or equi-biaxially stretched layers, we have $\alpha_1=\alpha_2$ and they become negative together for $E>E_c$, meaning that polar symmetric solutions of the Bessel type become possible immediately above the critical voltage, see Fig.\ref{Energy1}.
When the film is pre-stretched with $\lambda_1>\lambda_2$, for example, the first coefficient that becomes negative is $\alpha_2$, meaning that sinusoidal solutions in the direction of least stretch become possible, see Fig.\ref{Energy2}.

With this work we improved the current understanding of the catastrophic nature of electro-elastic instabilities in thin films. 
This unsolved problem goes back much further than the recent technological interest in applications of soft dielectrics. 
As early as 1880, R\"ontgen \cite{Roentgen} recorded the possibility of stretching natural rubber using sprayed-on electric charges. 
In 1955, Stark and Garton \cite{StarkGarton} detected a previously unrecognised form of breakdown, due to mechanical deformation of electrically irradiated thin films of cross-linked polythane. 
The first experimental evidence that dielectric breakdown is due to strong thinning localisation is the work of Blok and LeGrand \cite{BlokLeGrand} in 1969. 
They were the first to provide clear optical evidence (Fig.\ref{BlokLeGrand}) of highly localised thinning in voltage-controlled polymer films, which brought them to speculate that {\it it is experimentally impossible to deform the entire area of the dielectric without using an immense stress}, thus inferring the energetic convenience of localising deformations above a critical voltage. 
Remarkably, Blok and LeGrand's intuition remained largely unexplored to date. 
Here we further their intuition and find a completely new paradigm for the analysis of electromechanical instabilities in dielectric films, both one-side constrained and unconstrained.
The great majority of technological applications based on dielectric elastomers is dependent on whether wrinkling may or may not be anticipated by electromechanical instability, e.g. tunable adhesion, wetting, open-channel microfluidics, etc.\cite{YaKL10,CaHu12} and on-demand fluorescent patterning \cite{Wang14}.
By re-defining and furthering the concept of electromechanical instability of dielectric films, the paper essentially implies that new experimental campaigns and new analytical studies based on formula \eqref{EgenEcr} are now required to generate a finer physical picture of the catastrophic thinning phenomenon.

\clearpage

\begin{center}
{\huge \textbf{Supplementary material}}
\end{center}


\subsection*{Model derivation} 


Thermodynamical considerations for electro-elastic materials undergoing {relatively slow} motions lead to a formulation of electro-elastic equilibrium as a minimisation problem for a suitable {\it total free energy} functional \cite{Fosdick,Hong,Miehe1}. 
The undeformed configuration of the electro-elastic film is a flat prismatic plate $R$ with uniform cross section and small thickness $h$. 
The lower electrode is assumed to be at zero voltage and the upper electrode at voltage $V$. 
Our analysis is valid for `ideal dielectric elastomers' \cite{ZhSu07}, for which the polarisation is fluid-like and  independent of the deformation. 
Under these assumptions, when voltage is fixed the total free energy  takes the form 
\beq\label{energy1}
\hat\Psi({\bf f},\tilde{\bf d}) = U({\bf f}) - \frac{V\,Q({\bf f},\tilde{\bf d})}{2} - \scrW({\bf f}), 
\eeq
where $U$ is the free elastic energy, ${\bf f}$ the film deformation, $\tilde{\bf d}$ the (Lagrangian) electric displacement, $Q$ the total charge on the upper electrode and $\scrW$ the work of mechanical forces on the film edges. 

When the film thickness is small, we find approximated solutions $\tilde{\bf d}_{\bf f}$ to the electro-static problem for arbitrary deformations, and the total free energy can be recast as
\beq\label{energy2}
\Psi({\bf f}) = \hat\Psi({\bf f},\tilde{\bf d}_{\bf f}) = U({\bf f}) - \frac{Q({\bf f},\tilde{\bf d}_{\bf f})\,V}{2} - \scrW({\bf f}).
\eeq
Now the problem can be formulated in the familiar framework of classical elasticity, where deformation is the only unknown. 
To capture inhomogeneous thinning we minimise the energy \eqref{energy2} over an approximated class of deformations \cite{Coleman} bringing a point $\bx\in R$ to the current position
\beq\label{defgen}
{\bf f}(\bx) = {\bf g}(x_1,x_2) + \lambda_3(x_1,x_2)x_3\bi_3,
\eeq 
where ${\bf g}$ describes the planar component of deformation, perpendicular to the thickness direction $\bi_3$, and where $\lambda_3$ is the thickness stretch. 
We have $-h/2 \leq x_3 \leq h/2$ for unconstrained films and $0 \leq x_3 \leq h$ for one-side constrained films. 
In both cases we denote by $S$ the surface $x_3=0$. 

The gradient of the deformation \eqref{defgen} is 
\beq
\F = \nabla{\bf f} = \G + \lambda_3\bi_3\otimes\bi_3 + x_3\bi_3\otimes\nabla\lambda_3, 
\eeq
where $\G=\nabla{\bf g}$ is the in-plane deformation tensor, which may be recast as $\G=\lambda_1\bi_1\otimes\bi_1 + \lambda_2\bi_2\otimes\bi_2$ with $\bi_1\perp\bi_2$ both perpendicular to $\bi_3$ and where $\lambda_{1}$, $\lambda_2$ are the planar principal stretches. 
Volume variations are measured by $J=\det\F=\lambda_1\lambda_2\lambda_3$, so that incompressibility requires $\lambda_1\lambda_2\lambda_3=1$. Observe that if the film were exactly incompressible and, as in the case of one-side constrained film, the principal stretches $\lambda_1$ and $\lambda_2$ are fixed before gluing the membrane to the rigid electrode, it would follow that $\lambda_3=(\lambda_1\lambda_2)^{-1}$ must be constant, and no further deformations may take place. 
To describe creasing deformations we relax this constraint by taking a strain energy density for slightly compressible materials, separating the volumetric from the deviatoric part as follows  \cite{Pence}, 
\beq\label{WW}
\bbar W= W\left(I/J^{-2/3} \right) + \kappa\,\Phi(J), 
\eeq
where $I=\text{tr}(\mathbf F^T \mathbf F)$, $\Phi(J)$ is such that $\Phi(1) = \Phi'(1)=0$, and $\kappa$ is the {\it infinitesimal bulk modulus}. 
Then exact incompressibility is the limit $J \to 1$ and $\kappa / \mu\rightarrow\infty$ where $\mu = W'(3)$ is the infinitesimal shear modulus.
The free energy stored in the plate can be calculated as $U=\int_S\int_b^c \bbar W(\F(\bx))\,dx_3\,dA$, where $dA=dx_1dx_2$. 

Regarding the electric part of the problem, the charge on the upper electrode is $Q=\int \s^+\,da^+$ where $\sigma^+$ is the charge density per unit area. 
Assuming that the external electric field is null, we have $\s^+=\e{\rm e}^+$ where $\e$ is the dielectric permittivity and ${\rm e}^+$ the normal component of the Eulerian electric field on the upper electrode. 
Thanks to the film slenderness \cite{PuglisiZurlo}, we can neglect curvature effects on the electric field and we consider that ${\rm e}^+ \simeq {\rm e} \simeq V/(h \lambda_3)$. 
This approximation is consistent with nonlocal effects for the mechanical behaviour of the electrodes  not having been taken into consideration \cite{Z2013,DPZ2014}. 
The surface stretch of the upper electrode is found from Nanson's formula as $da^+/dA = ||J\F^{-}\trasp\bi_3||_{x_3=c}$. 

The work term only needs to be expressed when dealing with unconstrained films (for one-side constrained films we will consider the planar stretches, if any, to be fixed.) 
Denoting by $s_{1}$, $s_2$ the Lagrangian edge line stress components (force per undeformed length) of the membrane, we find 
\beq
\scrW = h \int_S(s_1\lambda_1+s_2\lambda_2)\,dA.  
\eeq
Since the film is slender we can perform a Taylor expansion of the total free energy \eqref{energy2} in powers of $h$, truncating $o(h^3)$ terms. 
After through-thickness integration, we thus obtain that $\Psi({\bf f}) = \int_S \psi(\F(\bx))\,dA$ where $\psi$ is the surface free energy. 
Finally, we introduce the dimensionless electric field, 
\begin{equation}
E = ({V}/{h})\sqrt{{\e}/{\mu}},
\end{equation}
where $\mu$ is the {\it infinitesimal shear modulus}. 

For {\it one-side constrained} films, we consider $\lambda_1$, $\lambda_2$ to be homogeneous and prescribed by the pre-stretching of the film prior to attaching it to the rigid substrate. 
The film is not exactly incompressible, so $\lambda_3$ is the independent variable. 
We thus obtain \cite{ZOber}
\beq\label{energycreasing}
\psi(\lambda_3,\nabla\lambda_3) = h \ph(\lambda_3) + h^3 \left[\alpha_1(\lambda_3)\lambda_{3,1}^2 + \alpha_2(\lambda_3)\lambda_{3,2}^2\right],
\eeq
where $\lambda_{3,i}=\partial\lambda_3/\partial x_i$ and  
\beq\label{alfacreasing}
\begin{split}
& \ph = W(I/J^{2/3}) + \kappa\,\phi(J) - \mu E^2 j /(2\lambda_3),\\
& \alpha_1 = 4 J^{1/3} W'(I/J^{2/3}) - 3\mu E^2 \lambda_2^2,\\
& \alpha_2 = 4 J^{1/3} W'(I/J^{2/3}) - 3\mu E^2 \lambda_1^2.
\end{split} 
\eeq
Here,  $I = \lambda_1^2 + \lambda_2^2 + \lambda_3^{2}$ and $j=\lambda_1\lambda_2$ (which is fixed). 

\bigskip

For {\it unconstrained} films, we impose exact incompressibility ($\lambda_3=1/j$), and then 
\beq\label{energypullin}
\psi(\lambda_i,\nabla\lambda_3) = h \ph(\lambda_i) + h^3 \left[\alpha_1(\lambda_i)\lambda_{3,1}^2 + \alpha_2(\lambda_i)\lambda_{3,2}^2\right],
\eeq
where $i=1,2$ and
\beq\label{alfapullin}
\begin{split}
& \ph = W(I) - \mu E^2 j^2/2 - s_1\lambda_1 - s_2\lambda_2,\\
& \alpha_1 = \left[4 W'(I) - 3\mu E^2 \lambda_2^2\right]/48, \\
& \alpha_2 =  \left[4 W'(I) - 3\mu E^2 \lambda_2^1\right]/48.
\end{split} 
\eeq
Here,  $I=\lambda_1^2+\lambda_2^2+(\lambda_1\lambda_2)^{-2}$. 
Note that for equi-biaxially stretched films, $\lambda_1=\lambda_2=\lambda_3^{-1/2}=\lambda$, say.


\subsection*{Energy minimisation} 


Here we exploit known necessary conditions for the existence of minimisers from direct methods of calculus of variations \cite{Dacorogna}. 
Assessing these necessary conditions does not rely on finding explicit nontrivial solutions but rather on direct inspection of the energy. 

Consider a scalar function $u(\bx):S\subset\bbR^2\rightarrow\bbR$ with $u=u_0$ prescribed on the edge $\partial S$ and consider the energy functional
\beq\label{functional}
\Psi(u) = \int_S\psi(\bx,u(\bx),\nabla u(\bx))\,dA.
\eeq
According to direct methods of calculus of variations \cite{Dacorogna}, {\it minimisers} of $\Psi$ exist provided that: 
\begin{itemize}
\item[(h1)]  $\bxi\rightarrow\psi(\bullet,\bullet,\bxi)$ is convex,
\item[(h2)]  there exist $p>q\geq 1$ and $a_1>0$, $a_2, a_3\in\bbR$ such that $\psi(\bx,u,\bxi)\geq a_1||\bxi||^p + a_2||u||^q+ a_3$ for all $(\bx,u,\bxi)$. 
\end{itemize}
{\it Uniqueness} of minimizers of $I(u)$ holds provided that
\begin{itemize}
\item[(h3)] $(u,\bxi) \rightarrow \psi(\bullet,u,\bxi)$ is strictly convex
\end{itemize}
Condition (h3) implies (h1)-(h2). 
Candidate minimisers of $I(u)$ must be sought among the stationary solutions of the functional $\Psi(u)$, found from the Euler-Lagrange equations
\beq\label{ELgen}
\frac{\partial\psi}{\partial u} = \Div\left(\frac{\partial\psi}{\partial\nabla u}\right)\hspace{20pt} u=u_o\,\,\,\,\text{on}\,\,\,\,\partial S. 
\eeq
Our main achievement here is to show that electromechanical instability corresponds to violation of the condition (h1).

Observe that Hessian approach consist in checking the convexity of $\psi$ in $u$, which is not a necessary condition for the existence of minimisers but is rather related to condition (h3), hence to uniqueness. 


\subsection*{Analysis of creasing for one-side constrained films} 


Here the prescribed pre-streches $\lambda_1,\lambda_2$ are homogeneous and the only unknown is the function $\lambda_3$. 
Homogeneous configurations $\lambda_3^0$ are found by solving the algebraic equation $\partial\ph/\partial\lambda_3=0$. 
Note that or large $\kappa$ the material is almost incompressible and $\lambda_3^0 \simeq (\lambda_1\lambda_2)^{-1}$.

A necessary condition for $\lambda_3^0$ to be a minimiser of $\Psi(\lambda_3)$ is that (h1) holds. 
It is immediate to check that this condition is satisfied as long as the functions $\alpha_{1}$, $\alpha_2$ of \eqref{energycreasing}$_{2-3}$ are positive, that is as long as
\beq
E < E_c = \frac{2}{\sqrt{3}} \sqrt{\frac{W'(I/J^{2/3})}{\mu}}\frac{\lambda_3}{J^{5/6}}\min(\lambda_1,\lambda_2),
\eeq
where $I$, $J$ are evaluated at $\lambda_1, \lambda_2, \lambda_3^0$. 
In the exact incompressibility limit the critical electric field above tends  to
\beq\label{MainAppendix}
\boxed{
E_c = \frac{2}{\sqrt{3}} \sqrt{\frac{W'(I)}{\mu}}\min\left(\frac{1}{\lambda_1},\frac{1}{\lambda_2}\right),}
\eeq
which is our main finding. 
We show below that this threshold is the same for unconstrained incompressible films. 
If the function $\psi(\lambda_3,\nabla\lambda_3)$ is strictly convex in both arguments, then the homogeneous solution $\lambda_3^0$ is also unique under homogeneous boundary conditions, according to (h3). 

As soon as (h1) is violated the energy functional $\Psi(\lambda_3)$ no longer admits minimisers in the class of deformations \eqref{defgen}; in particular, homogeneous configurations cannot be energy minimisers. 
Observe that this conclusion holds even if (h1) is violated while $\psi(\lambda_3,\bullet)$ is still convex in $\lambda_3$, which means that our condition is more restrictive than the positivity of the Hessian condition. 
Clearly, violation of (h1) does not exclude that other inhomogeneous solutions may exist outside the class \eqref{defgen}; 
nonetheless, we can show that immediately above $E_c$ inhomogeneous {\it failure precursors} become possible within the class \eqref{defgen}, that can lower the energy below $\Psi(\lambda_3^0)$, although the energy density has no minima beyond this threshold. 
For this reason we call these failure precursors {\it catastrophic}. 

To find out what these inhomogeneous precursors look like at their onset and to shed light on the physics of the problem, we consider a specific form for the strain energy density, the compressible neo-Hookean material \cite{Pence} with $W(I)=\mu(I/J^{2/3} - 3)/2$ and $\phi(J)=(J^2+J^{-2}-2)/8$,  and we look for inhomogeneous solutions $\lambda_3(\bx) = \lambda_3^0 + w(\bx)$ with $w$ small and $\lambda_3^o\simeq 1/(\lambda_1\lambda_2)$. The equation for $w$ is then
\beq\label{HelmholtzCreases}
\lambda_2^2\left(E^2 - \frac{2}{3}\frac{1}{\lambda_2^2}\right) w_{,11} + \lambda_1^2\left(E^2 - \frac{2}{3}\frac{1}{\lambda_1^2}\right) w_{,22} + \frac{\Gamma}{h^2} w = 0,
\eeq
where $
\Gamma =\left[16 + \lambda_1^2\lambda_2^2\left(6\kappa/\mu - 4\lambda_1^2 - 4\lambda_2^2 + 5\lambda_1^2\lambda_2^2 E^2\right)\right]/3$. 
To find out whether non-trivial solutions are possible, first observe that for nearly incompressible materials ($\kappa/\mu \gg 1$), $\Gamma$ is positive for all values of pre-stretch. Furthermore, the coefficients of $w_{,11}$ and $w_{,22}$ remain negative as long as $E<E_c$, meaning that no bounded, non-trivial, real solutions exist below this threshold; this clearly agrees with our general findings. 
However, as soon as one (or both) of the coefficients becomes positive, nontrivial solutions exist; this happens precisely when (h1) is violated. 

In the absence of pre-stretch ($\lambda_1=\lambda_2=1$), we  find radial solutions of the form
\beq
 w(\varrho) = C J_0\left(\frac{\omega\varrho}{h}\right), \qquad \varrho=\sqrt{x_1^2+x_2^2},
\eeq
where $J_0$ is the Bessel function of the first kind, $\omega =\sqrt{\Gamma/(E^2 - 2/3)}$ and  $C$ is an undetermined amplitude coefficient. 
In the presence of pre-stretch, experiments show that periodic creases appear perpendicularly to the direction of highest pre-stretch. 
When $\lambda_1>\lambda_2$, say, as soon as the electric field falls within the range $\sqrt{2/3}/\lambda_1<E<\sqrt{2/3}/\lambda_2$, the coefficient of $w_{,22}$ becomes positive while the coefficient of $w_{,11}$ remains negative, meaning that periodic solutions of the form
\beq
w(x_1,x_2) = C_1\cos\left(\omega\,\frac{x_2}{h}\right) + C_2\sin\left(\omega\frac{x_2}{h}\right)
\eeq
become possible, where $\omega=\sqrt{\Gamma/(E^2 - E_c^2)}/\lambda_1$ and $C_1$, $C_2$ are constants. 
Here the mechanical interpretation is that periodic patterns will become possible in the direction of least stretch, whereas in equi-biaxially stretched dielectrics there is no preferential direction and polar symmetric solutions may be attained; this agrees with experiments. 

Observe that, for both types of precursors presented above, the period of the failure precursors tends to zero at the onset of instability and increases immediately above the critical voltage. 
It the context of pull-in instability analysis, it was shown that this effect may be regularised by taking into consideration non-local effects \cite{APL2013ns,IJNLMns,Z2013,DPZ2014,PuglisiZurlo}. 
However, observe that discussing the nature of the post-critical shape of creases is beyond the limits of our simplified treatment for precursors. 
A more refined analysis certainly requires a wider class of deformations \cite{Audoly}, together with higher-order effects  for both the electrical and mechanical behaviour of the electrode.


\subsection*{Calibration of the Arruda-Boyce model for creasing} 


We calibrated the  5-term Arruda and Boyce \cite{ArBo93} model to reproduce the experimental results reported in \cite{WangCreasingPrestretch} as follows. 
First we use the Arruda-Boyce truncated strain energy density
\begin{multline}
W(I) = \mu_0
\left[\frac{I - 3}{2} + \frac{I^2 - 9}{20\,n} + \frac{11 (I^3 - 27)}{1050  \,n^2}\right.
    \\  + \left.\frac{19(I^4 - 81)}{7000\,n^3} + \frac{519 (I^5-243)}{673750\,n^4}
\right],
\label{arruda}
\end{multline}
where $n$ is the number of links in the 8-chain model.
Note that in \eqref{arruda} $ \mu_0 \ne 2 W'(3)$, i.e. $\mu_0$ is not the infinitesimal shear modulus (in fact, $\mu = 2W'(3)= \mu_0[1+3/(5n) + 99/(175n^2) + 513/(875n^3 + 42039/(67375 n^4)]$).
We then specialise the general equation \eqref{MainAppendix} to a uniaxial pre-stretch as described in \cite{WangCreasingPrestretch}, so that $\lambda_1=\lambda$, $\lambda_2=\lambda_3=\lambda^{-1/2}$, and $I=\lambda^2+2\lambda^{-1}$. 
When $\lambda=1$, Wang et al.  \cite{WangCreasingPrestretch} gave $E_c=1$ instead of their measured \cite{WaEE11} value $E_c=0.85$, meaning that they normalised their critical electric field--stretch curve with respect to that value. 
To model their data we thus have to calibrate our formula \eqref{MainAppendix} with respect to its first value $E_c =\sqrt{2/3} = 0.816$ at $\lambda=1$.
Then we show the curves found when the parameter $n$ varies from 2 to 7, indeed a realistic range for silicone materials used in electro-patterning applications, see Fig.4 in the main paper.


\subsection*{Analysis of pull-in for  unconstrained films} 


The analysis here follows  the same path as for creasing instability. 
Confining attention to equi-biaxially pre-stretched configurations under the action of equal  edge dead-loads, so that $\lambda_1=\lambda_2=\lambda_3^{-1/2}=\lambda$ and $s_1=s_2=s$, we obtain the general solution for the homogeneous loading branches by solving $\partial\ph/\partial\lambda=0$ as
\beq\label{homprestretch}
E^0(\lambda) = \sqrt{\frac{2 (\lambda^{-2} - \lambda^{-8}) W'(I) - s \lambda^{-3}}{\mu}},
\eeq
where $I = 2\lambda^2+\lambda^{-4}$. It is straightforward to check that also in this case the necessary condition for the existence of minimisers coincides with \eqref{MainAppendix}. 
Its specialisation to the equi-biaxial case is trivial:
\beq
E_c = \frac{2}{\sqrt{3}} \sqrt{\frac{W'(I)}{\mu}} \frac{1}{\lambda}.
\eeq
For the analysis of  failure precursors, the same treatment as that for creasing holds, and we do not repeat it for brevity. 
In this case the critical points are determined by the intersections of the $E_0$ curve with the $E_c$ curve. 
These expressions are used in the main paper (Fig.5) to capture the data of Wang et al. \cite{SuoGiant} (see below for the calibration of the model). 
\begin{figure}[!th]
\includegraphics[scale=0.5]{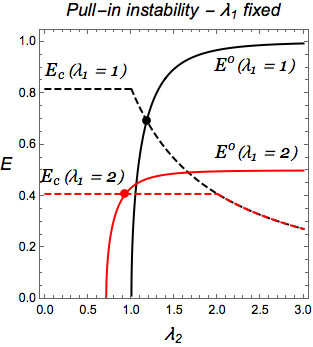}
\caption{\label{uniaxpullin}\footnotesize{Voltage versus $\lambda_2$ with $\lambda_1$ fixed. Intersection of the solid and dashed curves gives the upper voltage thresholds in both cases.}}
\end{figure}

Our formula also works when the extension is not equi-biaxial extension, a case where the Hessian approach also reveals its limitations. 
Take a neo-Hookean dielectric film where $\lambda_1>1$ is fixed and a traction $s_2$ is applied in direction $\bi_2$. 
Here the homogeneous configuration is retrieved from $\partial\ph/\partial\lambda_2=0$, giving
\beq\label{hompulluni}
E^0(\lambda_2) = (\lambda_1\lambda_2)^{-2}\sqrt{\lambda_1^2\lambda_2^3(\lambda_2-s_2/\mu) - 1}.
\eeq
We then find that along this homogenous loading curve, $\ph''(\lambda_2) = 1 + 3\lambda_1^{-2}\lambda_2^{-4} - (E^0)^2\lambda_1^2\geq 0$, with equality holding only for $\lambda_2\rightarrow\infty$. 
Hence, according to the Hessian approach, failure occurs at infinite stretch, which is clearly unrealistic. 
On the other hand, formula \eqref{MainAppendix} gives values of $E_c(\lambda_2)$ for each value of $\lambda_1$, represented by dashed curves in Fig.\ref{uniaxpullin}. 
The intersection of that curve with the curve of $E^0$ then gives the critical voltage for any pre-stretch.


\subsection*{Model calibration for pull-in} 


Tuning of the constitutive model with the experimental data provided in \cite{SuoGiant} was done directly on the lower frequency voltage-stretch curves, where viscoelastic effects are less relevant. 
The following Worm-Like Chain modification of the Gent model \cite{SaccoGent} 
\begin{multline}
W(I) =  \mu\frac{1+b}{1+2 b}J_m \ln\left(\frac{b}{1+b}\frac{I - 3}{J_m - (I - 3)}\right.\\
 +\left.\frac{J_m}{J_m-(I - 3)}\right),
\end{multline}
was used to obtain a best fit of the homogeneous loading curves. 
Here the best simultaneous fit for the four homogeneous loading curves was obtained with $b=-5$, $J_m=65$ and $\mu=20.5\,$ kPa. 




\end{document}